\documentclass[aps,pre,epsfigure,twocolumn]{revtex4-1}
\usepackage[colorlinks=true,linkcolor=blue,urlcolor=blue,citecolor=blue,pdfusetitle]{hyperref}
\usepackage[utf8]{inputenc}
\usepackage[english]{babel}
\usepackage{amsmath}
\usepackage[caption = false]{subfig}
\usepackage{graphicx,epstopdf}
\usepackage{blindtext}
\usepackage{lipsum}
\usepackage{amsfonts}
\usepackage{bbm}
\usepackage{amssymb}
\usepackage{enumerate}
\usepackage{color}
\usepackage{latexsym}
\usepackage{physics}
\usepackage{times,txfonts}

\newcommand{\Ccal}{\mathcal{C}}

\newcommand{\Ecal}{\mathcal{E}}

\newcommand{\Pcal}{\mathcal{P}}
\newcommand{\Qcal}{\mathcal{Q}}

\newcommand{\Tcal}{\mathcal{T}}
\newcommand{\Ucal}{\mathcal{U}}

\newcommand{\kets}[1]{| #1 \rangle}

\newcommand{\interpro}[2]{\langle #1 | #2 \rangle}

\newcommand{\trs}[1]{\text{Tr}#1}

\begin{document}

\title{Entanglement, coherence and charging process of quantum batteries}

\author{F. H. Kamin}
\affiliation{Department of Physics, University of Kurdistan, P.O.Box 66177-15175 , Sanandaj, Iran}

\author{F. T. Tabesh}
\email{f.tabesh@uok.ac.ir}
\affiliation{Department of Physics, University of Kurdistan, P.O.Box 66177-15175 , Sanandaj, Iran}

\author{S. Salimi}
\email{ShSalimi@uok.ac.ir}
\affiliation{Department of Physics, University of Kurdistan, P.O.Box 66177-15175 , Sanandaj, Iran}


\author{Alan C. Santos}
\email{ac\_santos@df.ufscar.br}
\affiliation{Departamento de Física, Universidade Federal de São Carlos, Rodovia Washington Luís, km 235 - SP-310, 13565-905 São Carlos, SP, Brazil}


\begin{abstract}
Quantum devices are systems that can explore quantum phenomena, like entanglement or coherence, for example, to provide some enhancement performance concerning their classical counterparts. In particular, quantum batteries are devices that use entanglement as the main element in its high performance in the powerful charging. In this paper, we explore the quantum battery performance and its relationship with the amount of entanglement that arises during the charging process. By using a general approach to a two and three-cell battery, our results suggest that entanglement is not the main resource to quantum batteries, where there is a non-trivial correlation-coherence trade-off as a resource for the high efficiency of such quantum devices.
\end{abstract}

\maketitle

Recently, the idea of quantum batteries $(QBs)$ has been proposed to exploit quantum effects in order to gain the charging time and charging power compared to their classical counterparts. The concept of quantum batteries was first introduced as two-level systems for energy storage and transmission to consumer centers~\cite{Alicki:13}. Therefore, the issue of efficient and operational quantum batteries is always an essential subject. In most scenarios, quantum batteries are considered as $N$ independent systems that are charged by a temporary field. However, so far there have been many efforts to model protocols to extract more work from a quantum battery, in particular by employing quantum entanglement~\cite{Binder:15,PRL2017Binder,Ferraro:18,Santos:19-c}.

As a new approach, the concept of quantum batteries is developed as many-body systems, where $N$ cells of a QB are charged locally \cite{Le:18,Andolina:19-2}, different than previous processes where the cells are jointly charged by using global operations. In this model, the quantum battery is presented as a one-dimensional Heisenberg spin chain composed of $N$ spins, which provides the intrinsic interactions between the spins and the possibility of entanglement. In a spin chain, we can consider a coupling given by the XXZ Heisenberg model, where an anisotropic parameter $\Delta$ develops a role in the dynamics of such a system. It is known that the XXZ Heisenberg chain has been applied to quantum batteries~\cite{Le:18}, but the role of the quantum correlations, e.g. entanglement and coherence, is yet an open question. Moreover, since it has been shown that entanglement is not necessary to optimal work extraction~\cite{PRL2013Huber}, this leads us to ask whether the quantum supremacy of QB is due to the entanglement.

To address this question, one needs to consider a suitable approach where the collective charging process can be done without entanglement generation. In this paper, we consider a two-qubit QB (a two-qubit cell), where we display the battery charge dynamics for both collective and non-collective (parallel) charging processes. Our results suggest that entanglement is not always the best resource to charge QBs, where in this scenario the coherence generation is the quantum resource for optimal charging of QBs. To end, we investigate the relation between
entanglement and coherence with the performance of three-qubit QB.

\emph{Ergotropy and charging process of quantum batteries.} The work extraction from quantum batteries is well defined by the \textit{ergotropy}~\cite{Allahverdyan:04}, where we can define the notion of \textit{passive states}, which are states where no amount of work can be extracted from them by unitary transformations. It is important to highlight the non-uniqueness of the passive states, in general~\cite{Perarnau-Llobe:15}. However, for pure states, the passive state can be well defined as the ground state of the system because it is the lowest energy state of the system~\cite{Santos:19-a}. Here, we focus on processes where the system is thermally isolated so that no heat is exchanged at any point during the process. We also consider cyclic processes, in the sense that the driving Hamiltonian is the same at the beginning and at the end of the dynamics. Since the system is thermally isolated, the evolution of state $\rho$ can be described by a unitary operator. Therefore, the extracted work is given by 
\begin{align}
\Ecal = W_{\text{max}} = \trs(\rho H_{0})-\max_{U\in\Ucal}\trs(U\rho U^{\dagger}H_{0}) \text{ , } \label{ErgGene}
\end{align}
where $\Ucal$ is the set of all accessible unitary evolution, and the internal (time-independent) Hamiltonian $H_{0}$ of the system can be decomposed as $H_{0}\!=\!\sum_{i}\nolimits \varepsilon_{i}\vert \varepsilon_{i}\rangle\langle \varepsilon_{i}\vert$, with $\varepsilon_{i+1} \geq \varepsilon_{i}$. It is possible to show that the work can be extracted from a system if and only if the system is non-passive, where a passive system has the form $\sigma_{\rho}=\sum_{i}p_{i}\vert \varepsilon_{i}\rangle\langle \varepsilon_{i}\vert$, where $p_{i+1}\leq p_{i}$~\cite{Allahverdyan:04,Gianluca:17}. That is, passive states are diagonal in the energy basis and do not have population inversions. Then, any unitary acting on $\rho$ can only increase its energy; and hence no work can be extracted from it. It easily follows that given a pure state the passive state reads as $\sigma^{\text{pure}}_{\rho}\!=\!\rho_{\text{g}}\!=\!\vert \varepsilon_{f}\rangle\langle \varepsilon_{f}\vert$, with $\vert \varepsilon_{f}\rangle$ being the fundamental state of $H_{0}$~\cite{Santos:19-a,Santos:19-c}. Therefore, the available energy of a QB unitarily charged reads
\begin{align}\label{W}
\Ecal = W_{\text{max}} = \trs(\rho H_{0})-\trs(\rho_{\text{g}}H_{0}) \text{ . }
\end{align}

Throughout the analysis presented here, we are dealing with unitary processes, then the above equation corresponds to the internal energy variation of the system concerning the energy scale defined by $H$.

\begin{figure}[t] 
	\includegraphics[scale=0.5]{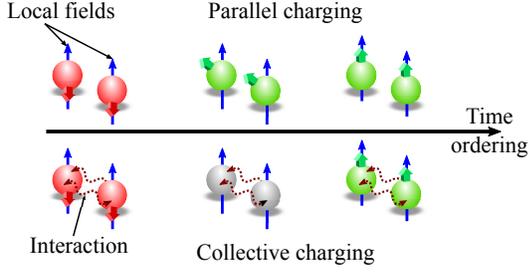}
	\caption{Schematic diagram of a two-cell QB, e.g. a two spins system, being charged through the parallel and collective charging, respectively. Local fields act on the cells and the cells interact with each other along a collective charging. In collective charging, the system can evolve through an entangled state (gray balls).}
	\label{fig1}
\end{figure}

\emph{Two-cell quantum batteries.} First, we start by introducing our physical model, as illustrated in Fig. \ref{fig1}, the two-qubit cell QB consisting of two coupled two-level systems. At the same time, in order to charge the QB, we need to consider that each cell couple individually with local fields. Without loss of generality~\cite{Santos:19-c}, we consider the driving Hamiltonian for our model in the interaction picture as $H\!=\!H_{\text{ch}}+H_{\text{int}}$, where $H_{\text{ch}}\!=\!\hbar \Omega\sum_{n=1}^{2}\sigma^{x}_{n}$, with $\sigma^{x}_{n}$ being the Pauli $X$-matrix acting on the $n$-th spin. The second Hamiltonian is the interaction one given by XXZ Heisenberg Hamiltonian given by
\begin{align}
H_{\text{int}}=J \hbar \left(\sigma^{x}_{1}\sigma^{x}_{2}+\sigma^{y}_{1}\sigma^{y}_{2}+\Delta\sigma^{z}_{1}\sigma^{z}_{2}\right) \text{ , } \label{Hint0}
\end{align}
where $\sigma^{i}~(i=x,y,z)$ are the Pauli matrices, $J$ is the strength of two-body interaction and $\Delta$ is a dimensionless parameter associated with the anisotropy of the chain. 

The status of the battery charging depends on the system state concerning the spectrum of the reference Hamiltonian $H_{0}$ considered here as $H_{0}\!=\!\hbar \omega_{0}\sum_{n=1}^{2}\sigma^{z}_{n}$, with identical Larmor frequency $\omega_{0}$ for both qubits. Here, as regard $\ket{\uparrow}$ and $\ket{\downarrow}$ are the ground and excited states of a single spin, respectively, we define the fully charged state of the battery as $\ket{\text{full}}\!=\!\ket{\uparrow\uparrow}$ with energy $E_{\text{full}}\!=\!2\hbar\omega_{0}$, and empty one as $\vert \text{emp}\rangle\!=\!\ket{\downarrow\downarrow}$  with low energy $E_{\text{emp}}\!=\!-2\hbar\omega_{0}$. Therefore, the maximum energy that can be stored in the battery reads $\Ecal_{\text{max}}\!=\!4\hbar\omega_{0}$.

Now, we investigate the charging process in two different situations. As sketched in Fig.~\ref{fig1}, we can drive the system with interaction between the cells (collective) and without interaction (parallel), where different results are expected~\cite{Binder:15,PRL2017Binder,Ferraro:18}. To study both processes we will start from the most general cases where interaction is considered. Since the Hamiltonian is time-independent, the system dynamics is given by
\begin{align}
\ket{\psi(t)} = \sum\nolimits_{n=1}^{4} c_{n}e^{-\frac{i}{\hbar} E_{n} t} \ket{E_{n}} \text{ , }
\end{align}
where $E_{n}$ are the eigenenergies of $H$ associated with the eigenstate $\ket{E_{n}}$ and $c_{n}$ are the coefficients of the expansion of the initial state of the system in the basis $\{\ket{E_{n}}\}$. The eigenenergies of $H$ are given by $E_{1}\!=\!J\Delta\hbar$, $E_{2} \!=\! -J(\Delta+2)\hbar$, $E_{3}\!=\!(J-\beta)\hbar$, $E_{4}\!=\!(J+\beta)\hbar$ with their respective eigenstates
\begin{align}
\vert E_{1}\rangle &= (\ket{\downarrow\downarrow} -\ket{\uparrow\uparrow})/\sqrt{2} \text{ , } ~
\vert E_{2}\rangle = (\ket{\downarrow\uparrow} -\ket{\uparrow\downarrow})/\sqrt{2} \text{ , } \nonumber\\
\vert E_{3}\rangle &=\gamma_{1}(\ket{\downarrow\downarrow} +\ket{\uparrow\uparrow})-\gamma_{2}(\ket{\downarrow\uparrow} +\ket{\uparrow\downarrow}) \text{ , } \nonumber\\
\vert E_{4}\rangle &=\gamma_{2}(\ket{\downarrow\downarrow} +\ket{\uparrow\uparrow})+\gamma_{1}(\ket{\downarrow\uparrow} +\ket{\uparrow\downarrow}) \text{ , }
\label{6a}
\end{align}
with
\begin{align}\label{7}
\gamma_{1} = \frac{2~\Omega}{\sqrt{2(\alpha+\beta)^{2}+8~\Omega^{2}}} \text{ , } ~~
\gamma_{2} = \frac{\alpha+\beta}{\sqrt{2(\alpha+\beta)^{2}+8~\Omega^{2}}} \text{ , }
\end{align}
where we defined $\beta \!=\!\sqrt{J^{2}(\Delta-1)^{2}+ 4~\Omega^{2}}$ and $\alpha \!=\!J(\Delta-1)$. 

\begin{figure*}[t!]
	\centering
	\subfloat[ ]{\includegraphics[scale=0.26]{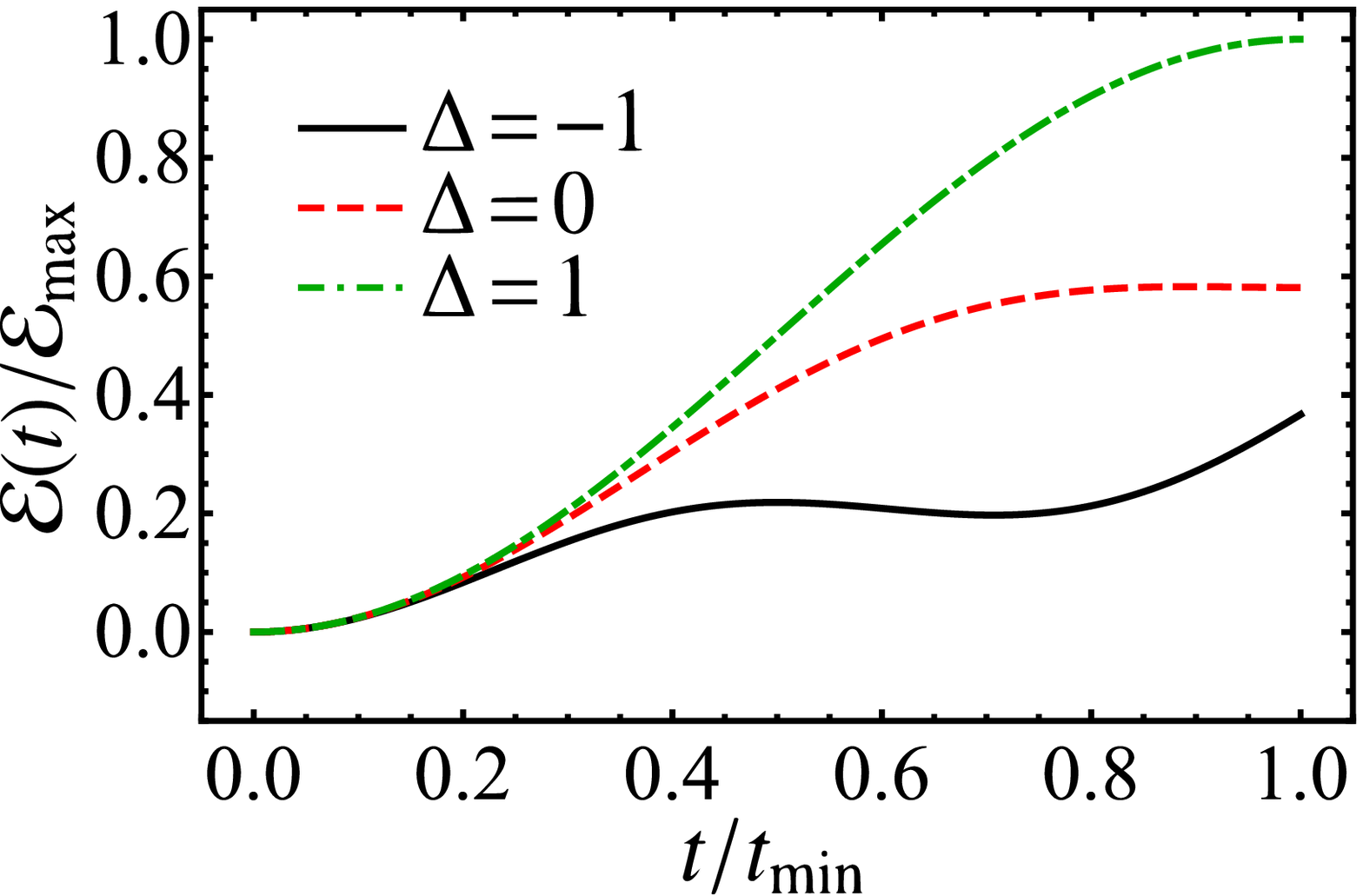}\label{Fig-Ergo}}~\subfloat[ ]{\includegraphics[scale=0.26]{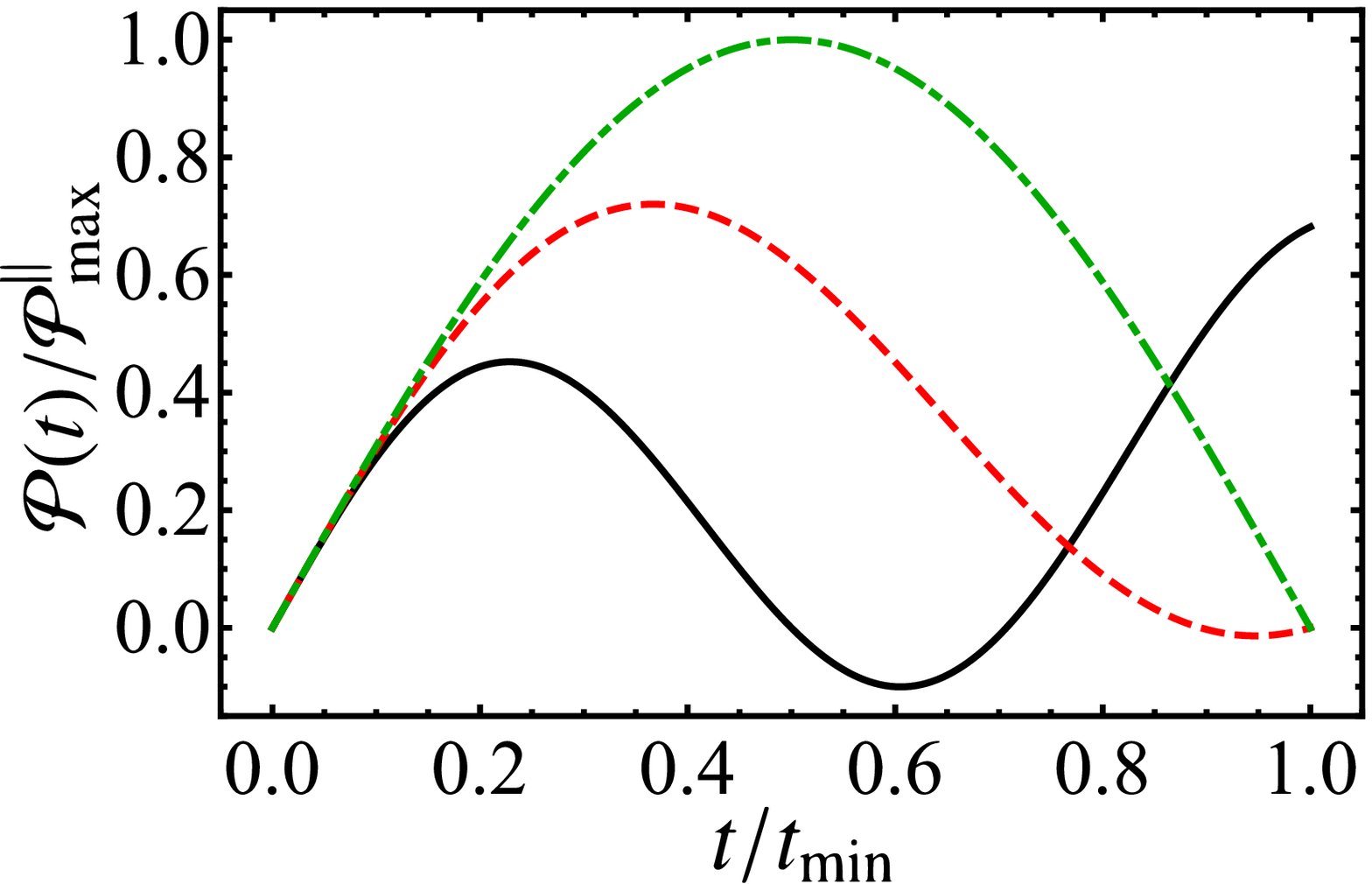}\label{Fig-Power}}~
	\subfloat[ ]{\includegraphics[scale=0.26]{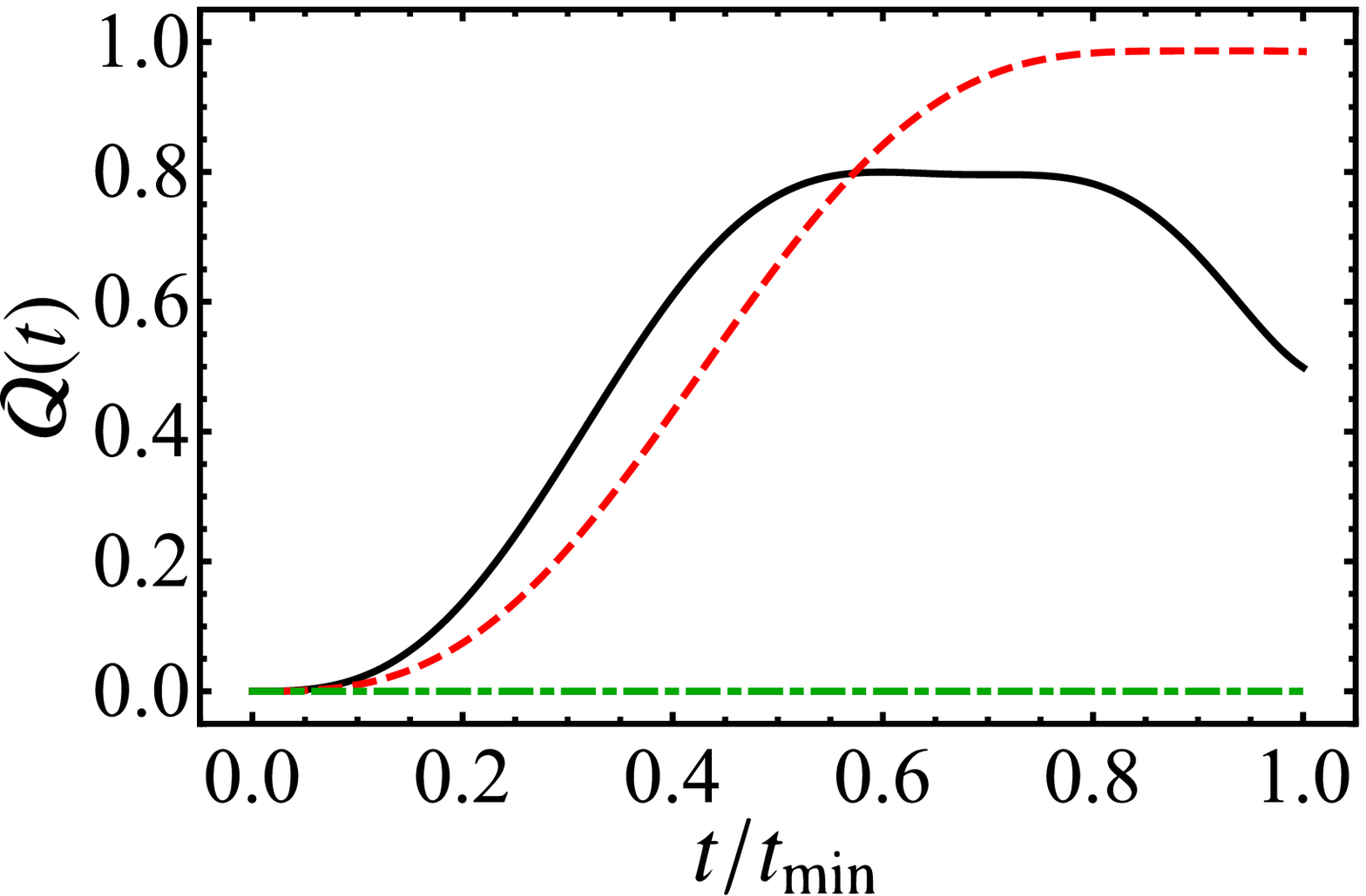}\label{Fig-Ent}}~\subfloat[ ]{\includegraphics[scale=0.26]{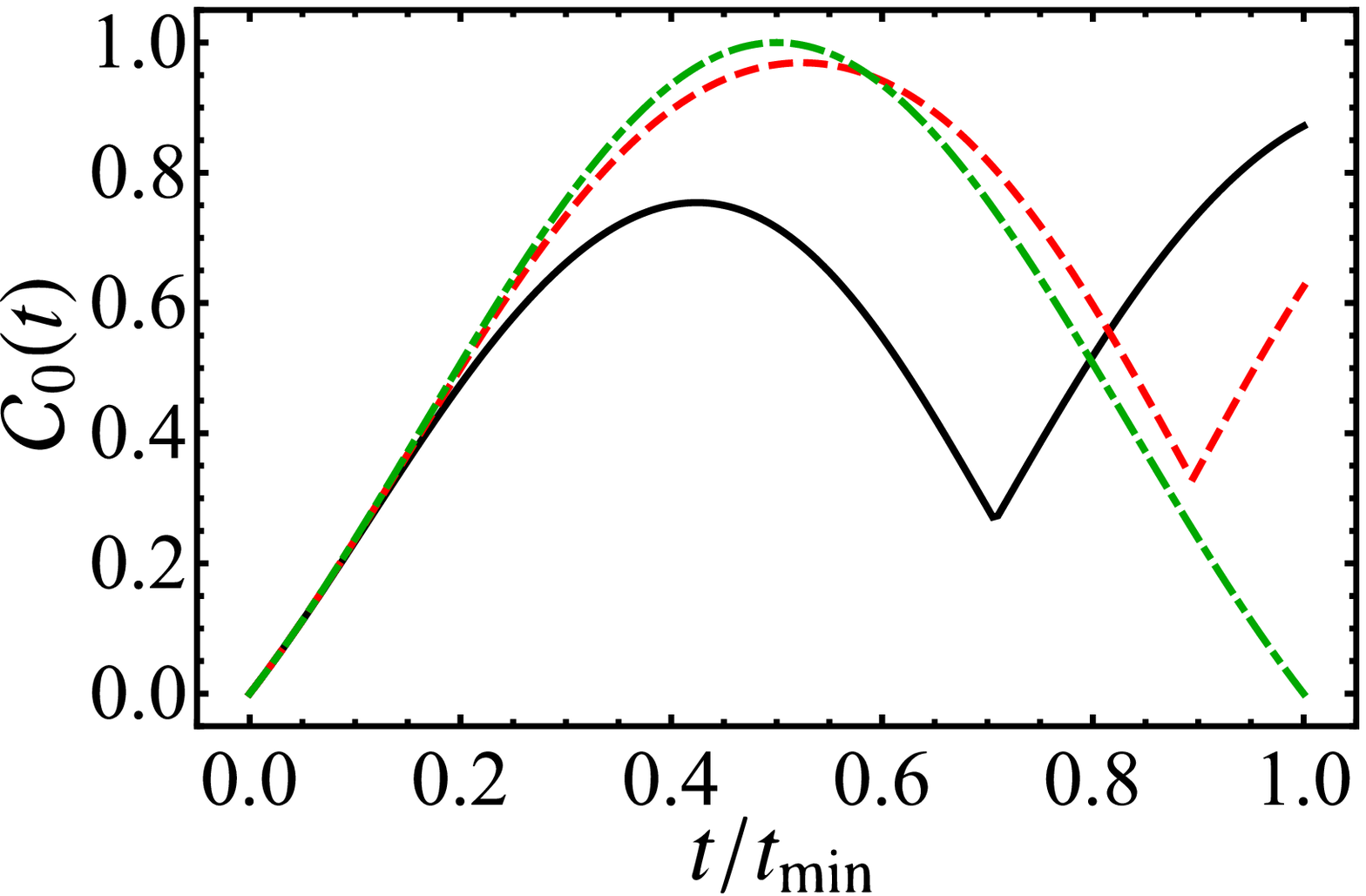}\label{Fig-Cohe}}
	\caption{Time evolution for~\eqref{Fig-Ergo} ergotropy,~\eqref{Fig-Power} instantaneous charging power,~\eqref{Fig-Ent} entanglement, and~\eqref{Fig-Cohe} coherence of the two-cell QB for different values of the anisotropy parameter $\Delta$. The coupling regime between the qubits is $J\!=\!\Omega$.}
	\label{Fig1}
\end{figure*}

As a first analysis, let us consider the parallel charging process of the battery ($J\!=\!0$), where each cell will independently evolves driven by the charging field. Therefore, from the above equations, we find the instantaneous ergotropy given by
\begin{align}
\mathcal{E}_{\parallel}(t) = \Ecal_{\text{max}} \sin^2(\Omega t) \text{ . } \label{Eparallel}
\end{align}

Immediately from this result, we establish the maximum average power for the parallel charging as $\bar{\Pcal}_{\text{max}}^{\parallel}\!=\!2\Ecal_{\text{max}}\Omega/\pi$, where we used that $t_{\text{min}}\!=\!\pi/2\Omega$ is the minimum time interval to get the maximum charge $\Ecal_{\text{max}}$. For the sake of completeness, from Eq.~\eqref{Eparallel}  we find the instantaneous power as
\begin{align}
\mathcal{P}_{\parallel}(t) = \frac{d \mathcal{E}_{\parallel}(t)}{dt} = \Pcal^{\parallel}_{\text{max}} \sin(2t\Omega) \text{ , }\label{Pparallel}
\end{align}
with $\Pcal^{\parallel}_{\text{max}}\!=\!\Ecal_{\text{max}}\Omega$ being the maximum instantaneous charging power. As we shall see, the quantities $\Pcal^{\parallel}_{\text{max}}$ and $\Ecal_{\text{max}}$ will be useful to study the role of the quantumness of the battery for a parallel and collective charging process.

\begin{figure}[t!]
	\centering
	\includegraphics[scale=0.3]{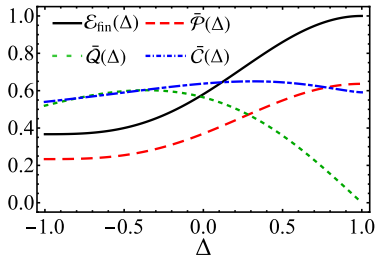}
	\caption{Graph for the quantities $\Ecal_{\text{fin}}(\Delta)$ (in unities of $\Ecal_{\text{max}}$), $\bar{\Pcal}(\Delta)$ (as multiple of $\Pcal_{\text{max}}^{\parallel}$), $\bar{\Qcal}(\Delta)$, and $\bar{\Ccal}(\Delta)$ as function of $\Delta$. The coupling regime between the qubits is $J\!=\!\Omega$.}
	\label{Fig-Delta}
\end{figure}

On the other side, the instantaneous ergotropy and charging power for the collective charging process ($J\!\neq\!0$) reads, respectively, as (See Appendix~\ref{ApA})
\begin{align}
\frac{\mathcal{E}_{\text{col}}(t)}{\Ecal_{\text{max}}}=\frac{1}{2}-\gamma^{2}_{1}\cos[(\beta+J\alpha)t]-\gamma^{2}_{2}\cos[(\beta-J\alpha)t]  \text{ ,} \label{Ecollective}
\end{align}
and
\begin{align}
\mathcal{P}_{\text{col}}(t) = 2\Pcal^{\parallel}_{\text{max}} \Omega\cos(\alpha Jt) \sin(\beta t)/\beta \text{ .} \label{Pcollective}
\end{align}

Now, as a first remark, we explore the role of the anisotropy parameter $\Delta$ in the special limit $\Delta\!\rightarrow\!1$, where we have $\alpha\!\rightarrow\!0$ and $\beta\!\rightarrow\! 2\Omega$, so the Eqs~\eqref{Ecollective} and~\eqref{Pcollective} give $
\mathcal{E}_{\text{col}}(t)|_{\Delta \rightarrow 1}\!=\!\mathcal{E}_{\parallel}(t)$ and $
\mathcal{P}_{\text{col}}(t)|_{\Delta \rightarrow 1}\!=\!\mathcal{P}_{\parallel}(t)$, recovering then results for the parallel charging process of a two-cell quantum battery. This quick remark allows us to conclude that the choice of $\Delta$ is relevant to the performance of QBs and lead us to ask: What is an effective collective charging process?
\begin{figure*}[t!]
	\centering
	\subfloat[ ]{\includegraphics[scale=0.19]{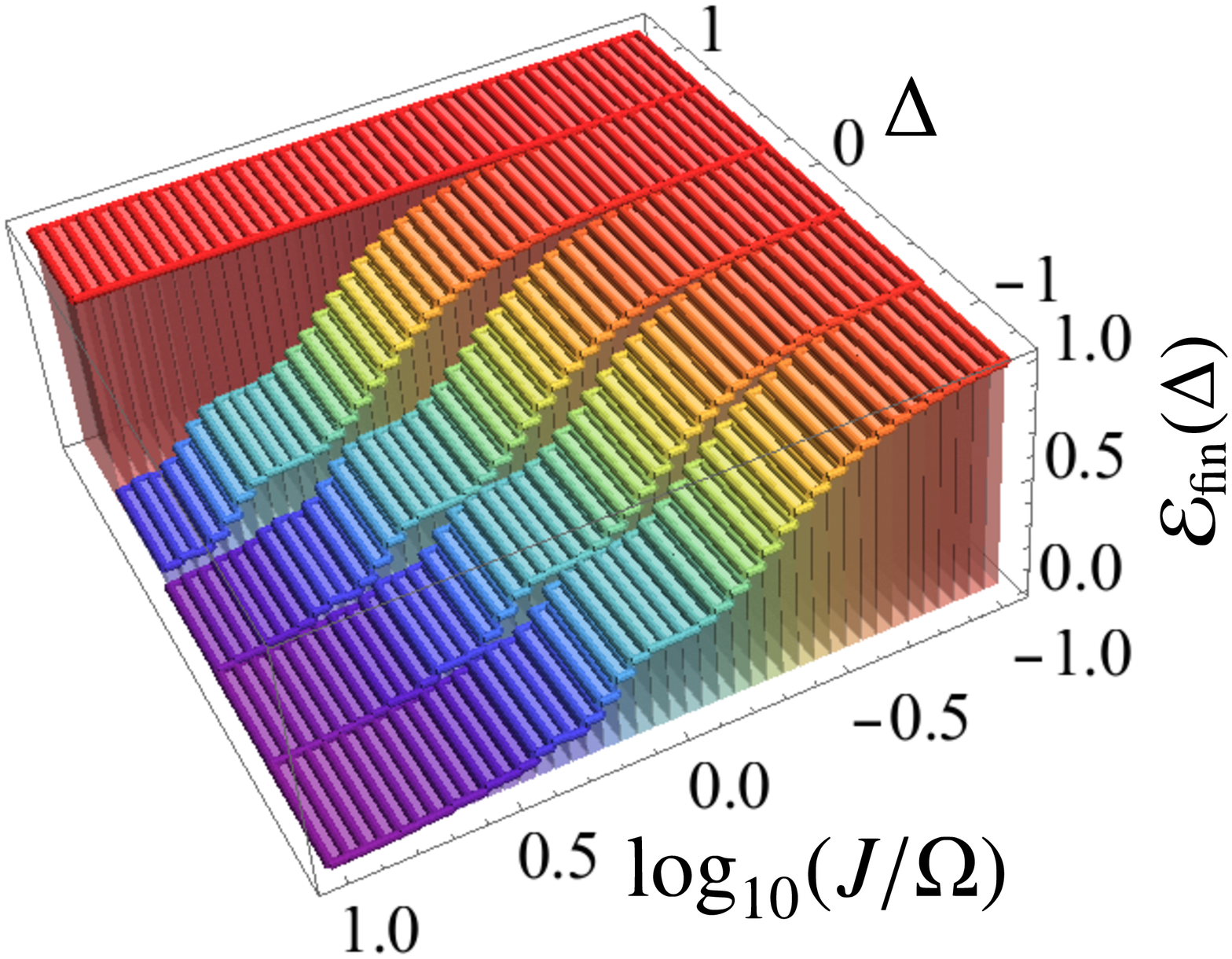}\label{Fig-3DErgo}}~\subfloat[ ]{\includegraphics[scale=0.19]{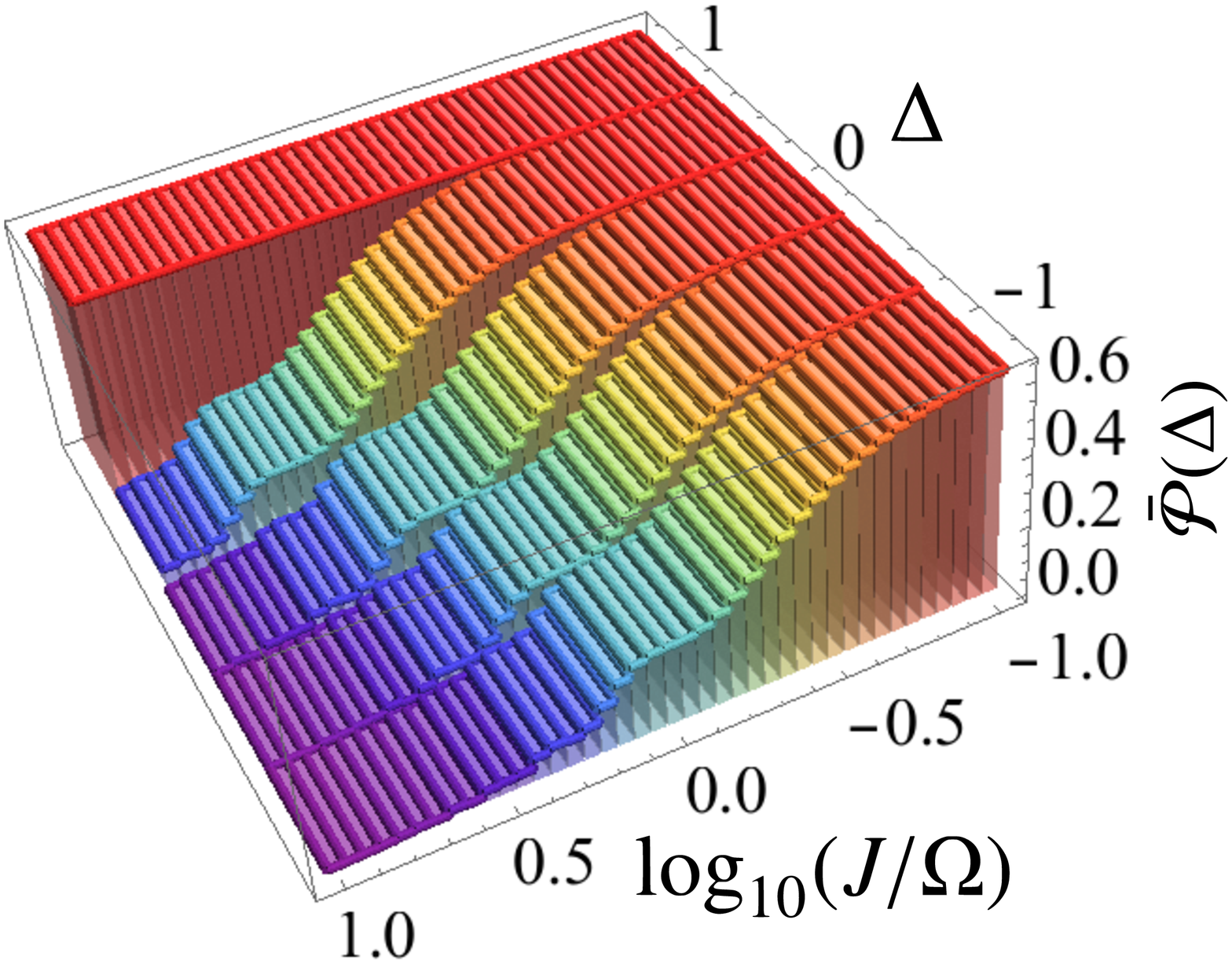}\label{Fig-3DPower}}~
	\subfloat[ ]{\includegraphics[scale=0.19]{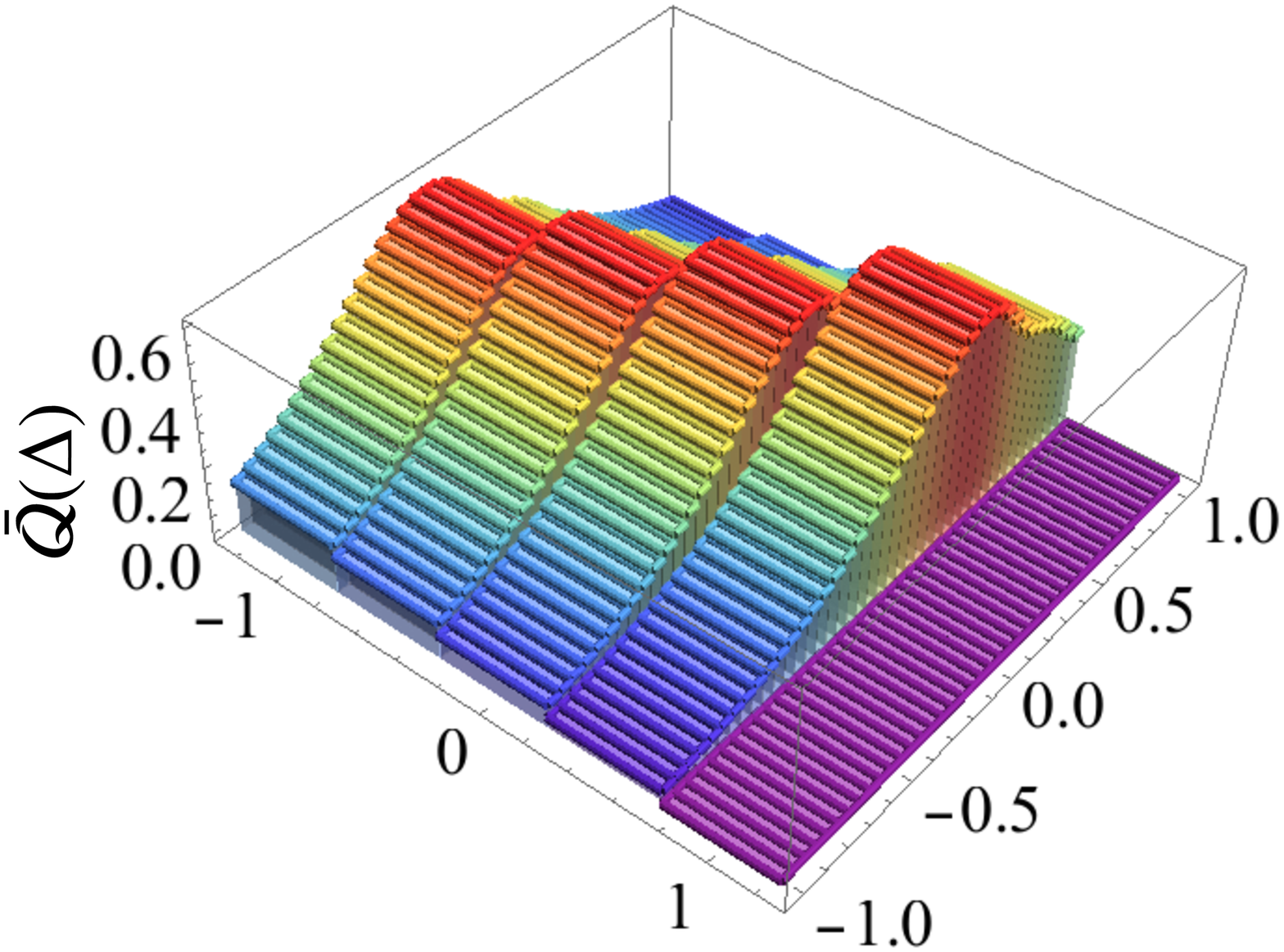}\label{Fig-3DEnt}}~\subfloat[ ]{\includegraphics[scale=0.19]{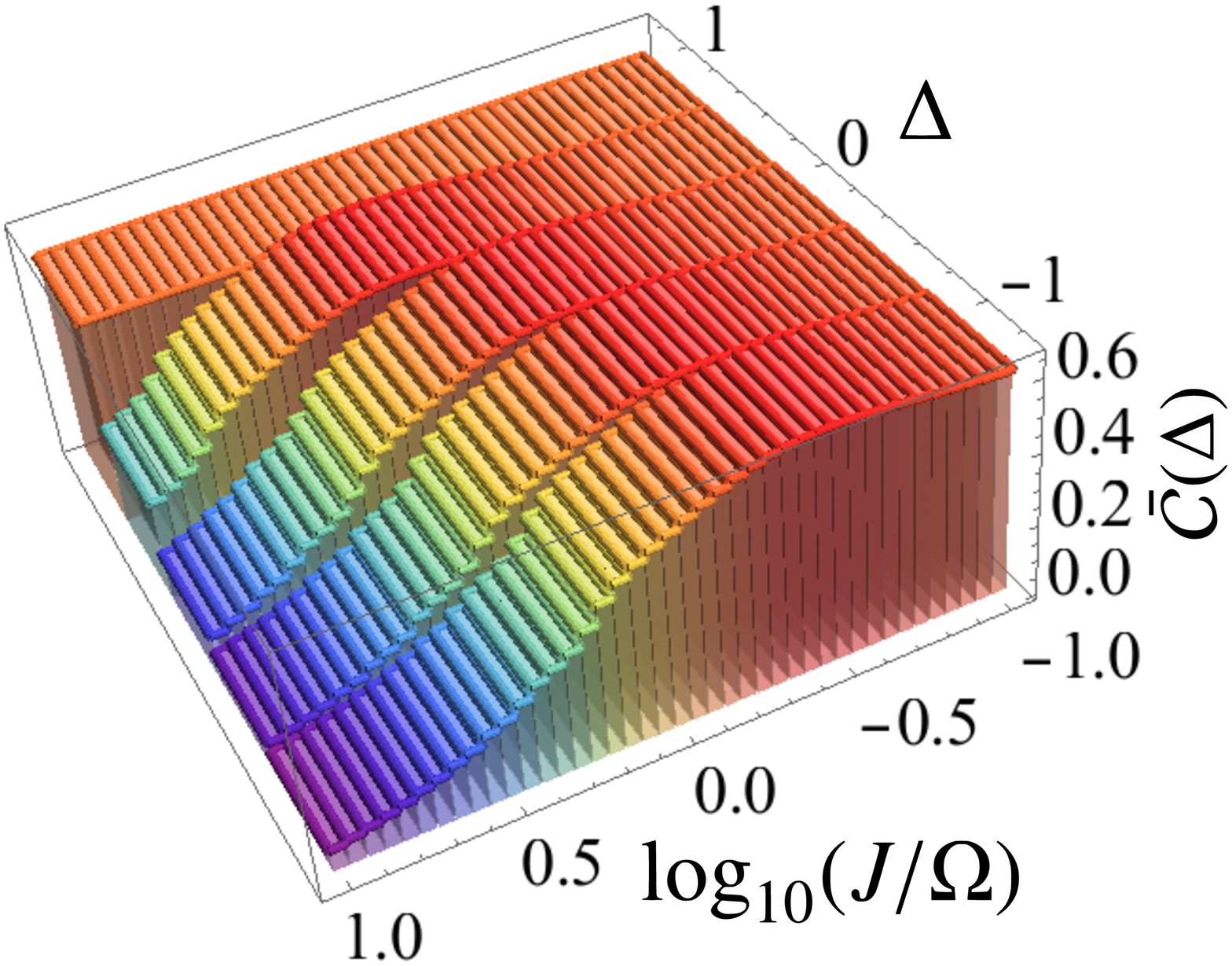}\label{Fig-3DCohe}}
	\caption{Graphs for the quantities $\Ecal_{\text{fin}}(\Delta)$ (in unities of $\Ecal_{\text{max}}$), $\bar{\Pcal}(\Delta)$ (as multiple of $\Pcal_{\text{max}}^{\parallel}$), $\bar{\Qcal}(\Delta)$, and $\bar{\Ccal}(\Delta)$ as function of $\Delta$ and $J/\Omega$ (in log scale). The regime of values for $J$ varies from $J=0.1\Omega$ ($\log_{10}(J/\Omega)\!=\!-1$) to $J=10\Omega$ ($\log_{10}(J/\Omega)\!=\!1$) and $\Delta\in\{-1,-0.5,0,0.5,1\}$.}
	\label{3DPlots}
\end{figure*}

\emph{Entanglement, coherence, and charging power.} The study of the quantumness of the two-cell QB will be addressed here from the amount of entanglement $\Qcal$ and normalized coherence $\Ccal_{\text{0}}$ of the system state. Given a pure state written in the reference basis as $\ket{\psi}\!=\!\alpha_{\uparrow\uparrow}\ket{\uparrow\uparrow}+\alpha_{\downarrow\uparrow}\ket{\downarrow\uparrow}+\alpha_{\uparrow\downarrow}\ket{\uparrow\downarrow}+\alpha_{\downarrow\downarrow}\ket{\downarrow\downarrow}$, we consider the entanglement given by the Wootters' measure of entanglement of a pair of qubits as~\cite{Wootters:98}
\begin{align}
\Qcal = 2|\alpha_{\uparrow\uparrow} \alpha_{\downarrow\downarrow} - \alpha_{\downarrow\uparrow} \alpha_{\uparrow\downarrow} | \text{ . } \label{Entanglement}
\end{align}
The energy content of the battery and the change in its energy distribution is of great interest in the context of quantum batteries. From this perspective, it is very useful to compute the coherence in the eigenstates of a bare Hamiltonian $H_{0}$ as the energy battery basis. As we shall see, this choice leads to a better understanding of the relationship between coherence as a quantum resource and the efficiency of quantum batteries. In addition, coherence in energy battery basis has been considered in recent works~\cite{Alexia:20,Baris:20}. Therefore, we define the coherence in the battery empty and charged basis as
\begin{align}
\Ccal_{\text{0}}(t) = (1/\Ccal_{\text{max}})\sum\nolimits_{i,j\neq i} |\rho_{ij}(t)| \text{ , } \label{Coherence}
\end{align}
with the quantity $\Ccal_{\text{max}}$ the maximum coherence of the system. For example, for a two-qubit state one reads $\Ccal_{\text{max}}\!=\!3$, which corresponds to the case $\kets{\psi_{\Ccal_{\text{max}}}}\!=\!(1/2)(\ket{\uparrow}+\ket{\downarrow})(\ket{\uparrow}+\ket{\downarrow})$. We define the above quantity by normalizing the definition of the $l_{1}$ norm of coherence~\cite{Baumgratz:14,Streltsov:17,Chitambar:19}, so that $0\!\leq\!\Ccal_{\text{0}}\!\leq\!1$. Then, from Eqs.~\eqref{Entanglement} and~\eqref{Coherence} one can study how much `quantum' the QB is. In addition, we are interested here in analyzing the role of entanglement for the charging process of the battery.

As previously discussed, through a parallel charging of the QB, the maximum charge state is achieved for minimum time interval $t_{\text{min}}$, then here we will analyze the dynamics of charging within the interval $t\!\in\!\Tcal_{\text{min}}\!=\![0,t_{\text{min}}]$. For our discussion, the time interval $\Tcal_{\text{min}}$ under consideration is appropriate, since we want to investigate both the role of correlations and the internal battery interaction. In this scenario, because $t_{\text{min}}$ is the minimum charging time of a parallel charging process, quantum correlations develops an important role if we can achieve the maximum charge for some time smaller than $t_{\text{min}}$. Otherwise, quantum correlations are not a resource. For completeness, it is worth mentioning that different values of $t_{\text{min}}$ has been considered in literature. For example, we can consider the minimum time given by the instant where we get maximum \textit{instantaneous} power $t_{\text{max}\Pcal}$~\cite{Le:18}. However, as we want to consider situations where we fully charge the battery, considering $t_{\text{max}\Pcal}$ as reference is not appropriated because $t_{\text{max}\Pcal}$ is not associated with the maximum charge instant. Actually, from definition of instantaneous power for the parallel charging (see Eq.~\eqref{Pparallel}), the instantaneous time $t_{\text{min}}$ of maximum values for $\Ecal(t)$ is associated with instantaneous power zero, because $t_{\text{min}}$ corresponds to a critical (maximum) point of $\Ecal(t)$ in time. Therefore, by considering the collective charging process $J\!\neq\!0$, Fig.~\ref{Fig1} shows the instantaneous power of the quantum battery for different choices of the anisotropy parameter $\Delta$. We highlight here the case with $\Delta\!=\!1$, in which no entanglement is present (as we can see in Fig.~\ref{Fig-Ent}) and the charging power is better than the other cases with $\Delta\!=\!0$ and $\Delta\!=\!-1$. However, such zero entanglement production does not mean the battery is classical. As we can see from Fig.~\ref{Fig-Cohe}, the maximum coherence is obtained in case where $\Delta\!=\!1$. Different from others works~\cite{Baris:20}, here we stress that the maximum ergotropy is not stored in the system coherence (the full charged state is $\ket{\uparrow\uparrow}$), but coherence works as a resource to speed up the charging process of the QB. It is worth mentioning that when we consider the case where effects that destroy coherence (decoherence process), the battery performance for the optimal configuration $\Delta\!=\!1$ becomes negatively affected and maximum charge is not achieved (see Appendix~\ref{BpB}). For this reason, we identify coherence as a resource to enhance the QB performance.

The role of the parameter $\Delta$ for the charging process can be better understood by defining average quantities for charge, power, entanglement, and coherence. We mean, one can define $\bar{\Pcal}(\Delta)$, $\bar{\Qcal}(\Delta)$, and $\bar{\Ccal}(\Delta)$ in the interval $t\!\in\![0,t_{\text{min}}]$, given by $\bar{X}(\Delta)\!=\!(1/t_{\text{min}}) \int_{0}^{t_{\text{min}}}X(t)dt$. In general, the average power is an important tool to investigate the charging performance of a quantum battery. However, in batteries where the spontaneous discharging is present~\cite{Santos:19-a,Santos:19-c} we can get ambiguous results, because the average power depends on the entire time window considered in the integration. For this reason, our analysis takes into account averaged values and the instantaneous quantities shown in Fig.~\ref{Fig1}, so that a robust analysis can be done~\footnote{In order to give an example of that, consider the time window as $\tau = 2t_{\text{min}}$. In this case we get $\bar{\Pcal}(\Delta)|_{\Delta=-1}\approx 0.3 \Pcal_{\text{max}}^{\parallel}$ and $\bar{\Pcal}(\Delta)|_{\Delta=1}= 0$, giving a result drastically different from that shown in Figs.~\ref{Fig1} and~\ref{Fig-Delta}.}. It is worth mentioning the physical meaning of $\bar{\Qcal}(\Delta)$ and $\bar{\Ccal}(\Delta)$, which can be understood as the average amount of entanglement and coherence, respectively, generated in the battery along the charging process. For completeness, we compute the value for the ergotropy at the end of the evolution $\Ecal_{\text{fin}}(\Delta)\!=\!\Ecal(t\!=\!t_{\text{min}})$, for different values of $\Delta$. From these sets of quantities, one can characterize the role of quantumness in the QB. In Fig.~\ref{Fig-Delta} we present the results for each quantify $\Ecal_{\text{fin}}(\Delta)$, $\bar{\Pcal}(\Delta)$, $\bar{\Qcal}(\Delta)$, and $\bar{\Ccal}(\Delta)$ as function of $\Delta$.

By remarking that the collective charging for the case where $\Delta\!=\!1$ is identical to the parallel charging process,  Fig.~\ref{Fig-Delta} suggests that entanglement-like quantum correlations in the QB are not beneficial for the performance of the QB considered in our study. It is indeed possible to see the difference of QB performance becomes enhanced for the situation in which the amount of entanglement generated along the entire evolution is vanishing. The quantum characteristic of the two-cell QB considered here is maintained due to the system state coherence, as we can see in Figs.~\ref{Fig1} and~\ref{Fig-Delta}. 

It is worthwhile to study the effect of parameter $J$ by creating different regimes to identify the optimal charging protocols for the QBs. Then, we compute the relevant quantities introduced in Fig.~\ref{Fig-Delta} as a two-variable function for $\Delta$ and the relative strength coupling $J/\Omega$, as shown in Fig.~\ref{3DPlots}. By comparing Figs.~\ref{3DPlots}, we remark situations in which by decreasing the coupling strength, the average work and power increase converging to the values given as in the region of $\Delta\!=\!1$, where the coherence plays an effective role in the charging process, becomes optimal in all situations for values of $J/\Omega$. Physically, it means whether we are increasing the pumping field intensity ($\Omega$), or we are just turning off the internal battery interactions. In both cases, we are close to the charging process with $\Delta\!=\!1$, which is independent on the strength coupling. Again, our regime of observation is given by $t\in [0,t_{\text{min}}]$, providing the optimal time window for our study. We remark that the behavior of the coherence in the system, Fig.~\eqref{Fig-3DCohe}, seems to be in agreement with the behavior of power and charge, Figs.~\eqref{Fig-3DErgo} and~\eqref{Fig-3DPower}, for all values of $J/\Omega$, while entanglement (average) behavior does not explain the increasing battery efficiency in the regimes considered here. It is worth highlighting that our results are consistent with the specific case of the strong-coupling limit $J\gg\Omega$ in reference \cite{Le:18}. Moreover, one can observe a significant reduction in the work and power of battery by increasing the coupling constant $J$ for other values of $\Delta\!\neq\!1$. In other words, these quantities tend to the maximum value at the limit $J\!\rightarrow\!0.1\Omega$. Consequently, this implies that non-zero anisotropy has no effect on the charging process of many-body quantum batteries in this regime. In fact, it means we have an intense charging field, then we expected that no internal interactions in the battery become relevant for the charging process. In the same way, in low intense regime of the charging field ($J\!\rightarrow\!10\Omega$), the dynamics is drastically governed by interaction and we can see the relevant role of the anisotropy in the battery charging performance.

\begin{figure*}[t!]
	\centering
	\subfloat[ ]{\includegraphics[scale=0.255]{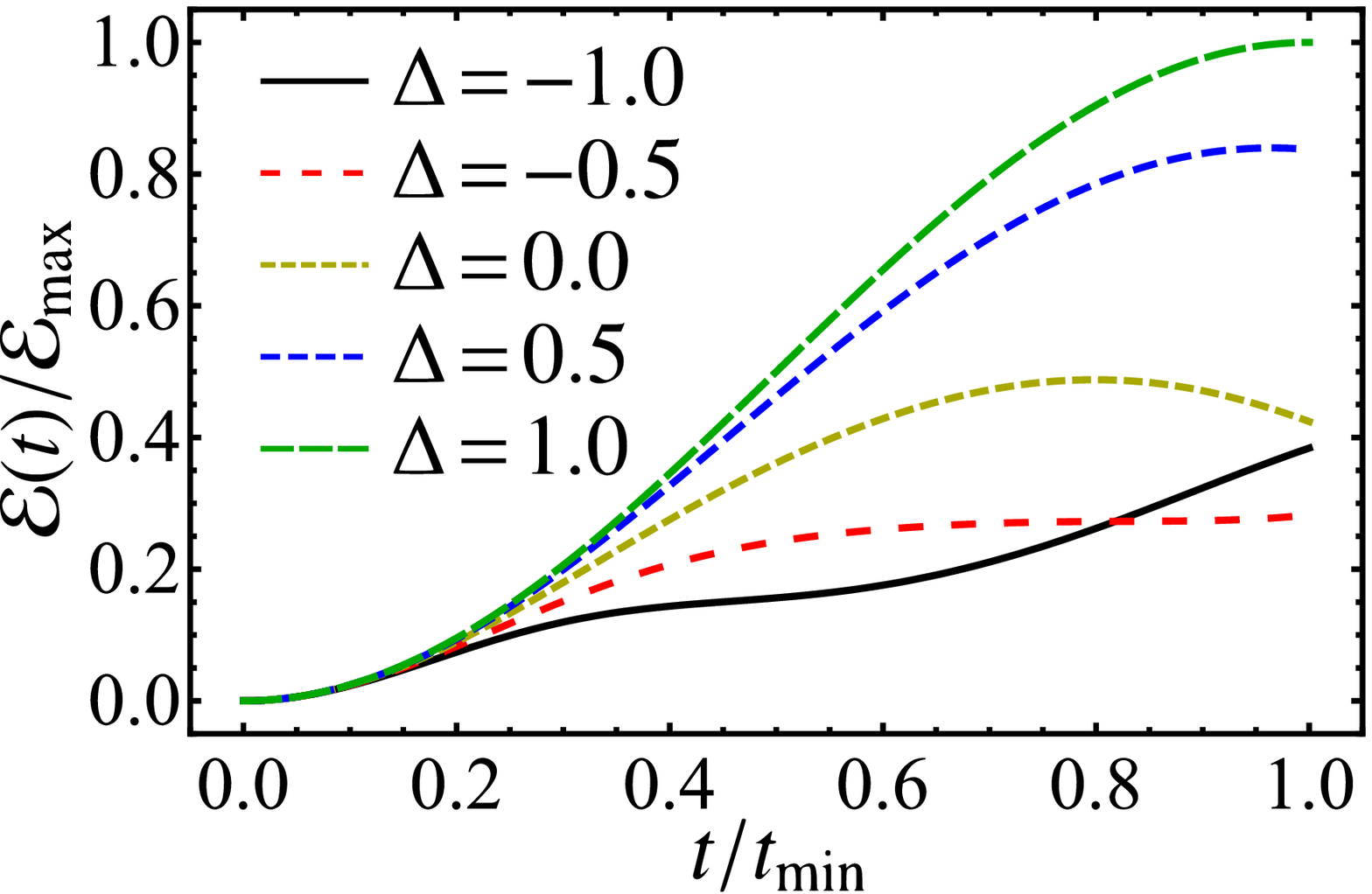}\label{Fig-Ergo2}}~\subfloat[ ]{\includegraphics[scale=0.255]{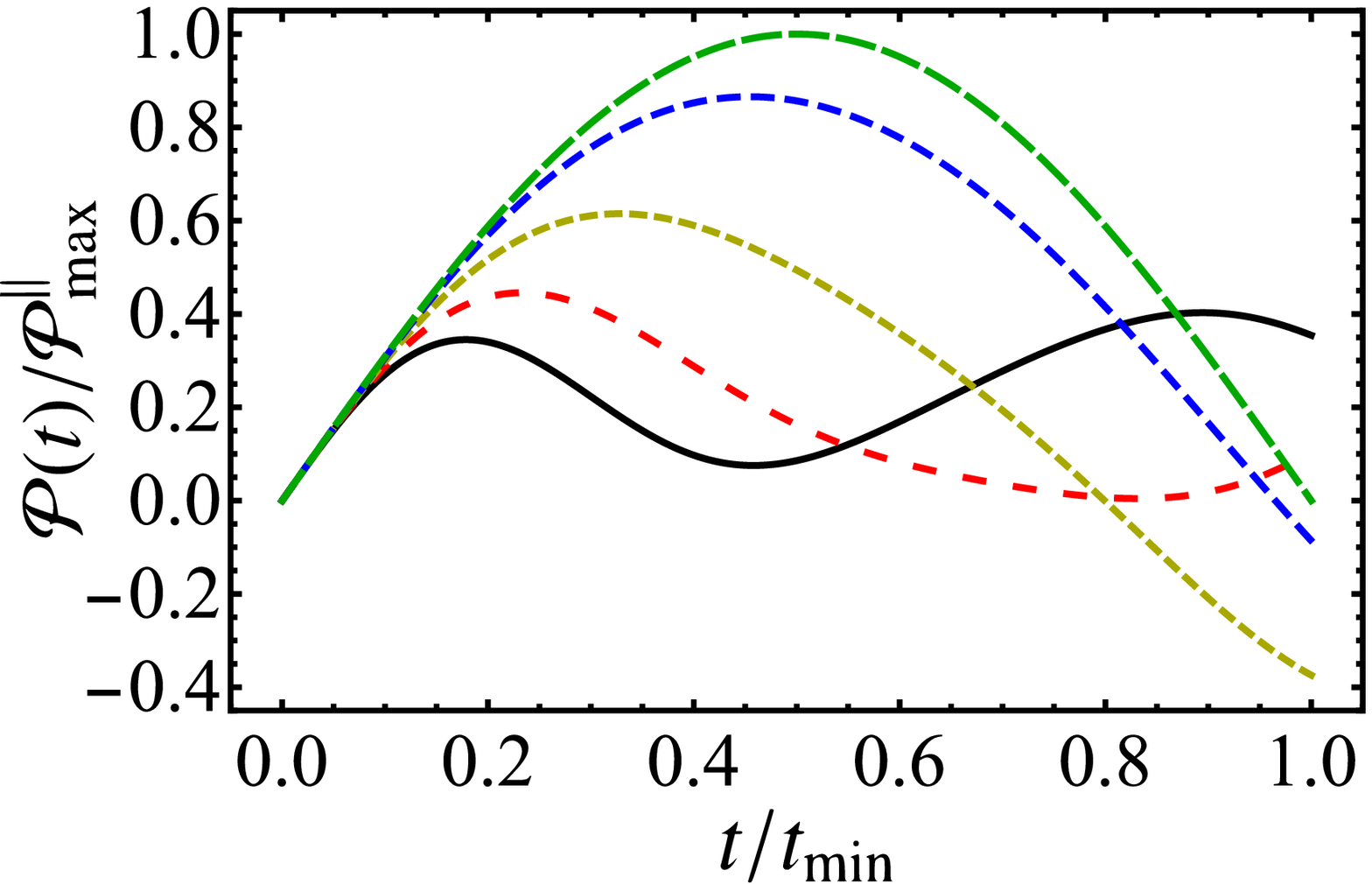}\label{Fig-Power2}}~\subfloat[ ]{\includegraphics[scale=0.255]{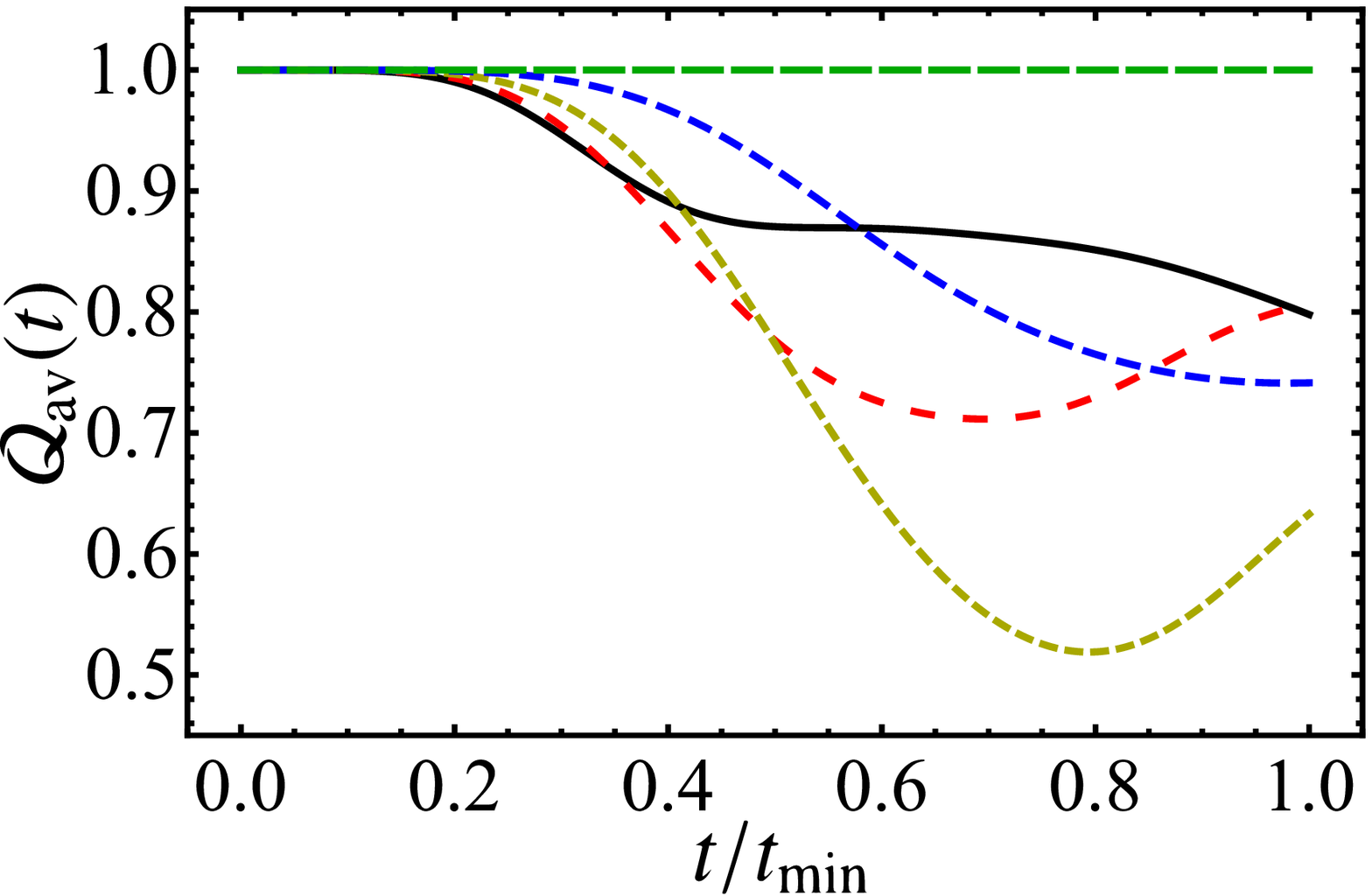}\label{Fig-Ent2}}~\subfloat[ ]{\includegraphics[scale=0.255]{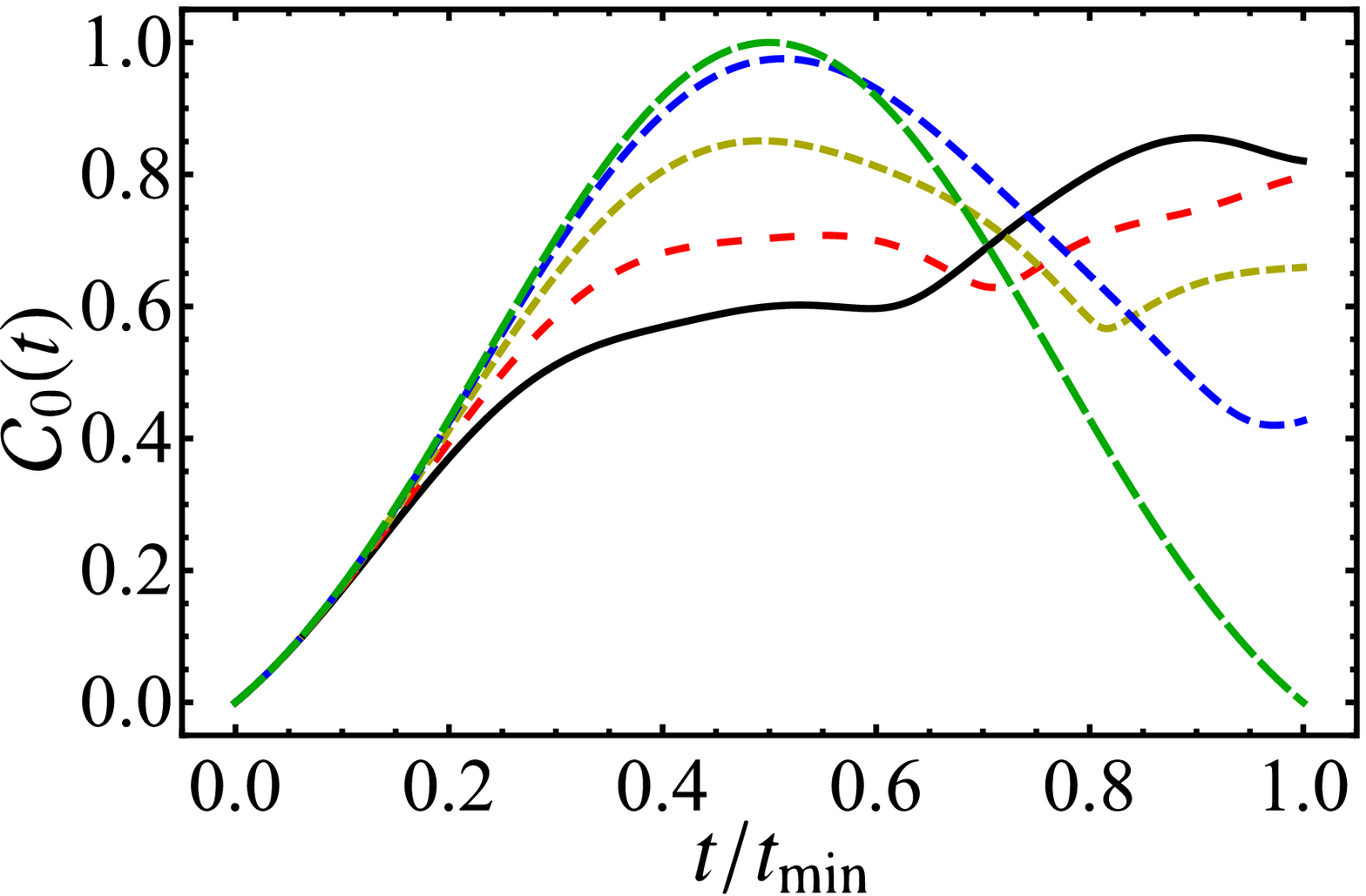}\label{Fig-Cohe2}}
	\caption{Time evolution for~\eqref{Fig-Ergo2} ergotropy,~\eqref{Fig-Power2} instantaneous charging power,~\eqref{Fig-Ent2} average entanglement and~\eqref{Fig-Cohe2} coherence of the three-cell QB for different values of the anisotropy parameter $\Delta$. The coupling regime between the qubits is $J\!=\!\Omega$, and for the three-cell QB we find $\Ecal_{\text{max}}\!=\!6\hbar\omega$ and $\Pcal_{\text{max}}^{\parallel}\!=\!\Ecal_{\text{max}}\Omega$.}
	\label{Fig4}
\end{figure*}

\emph{Three-cell QB.} As an immediate application, let us now discuss the quantumness of a three-cell QB. It can be done by adding a new cell to the battery, the new interaction Hamiltonian reads $H^{\prime}_{\text{int}}\!=\!J \hbar \sum_{n=1}^{2} (\sigma^{x}_{n}\sigma^{x}_{n+1}+\sigma^{y}_{n}\sigma^{y}_{n+1}+\Delta\sigma^{z}_{n}\sigma^{z}_{n+1} )$. Then, for this case we numerically solve the system dynamics $\rho(t)$ and compute the quantities $\Ecal(t)$ and $\Ccal_{0}(t)$ as done previously, but the quantities $\Pcal(t)$ and $\Qcal(t)$ need to be computed in a different way. Due to the numerical solution, to be practice, $\Pcal(t)$ is computed here from the energy current operator $\hat{\Pcal}$ as $\Pcal(t)\!=\!\trs(\hat{\Pcal}\rho(t))$, where~\cite{Santos:19-c}
\begin{align}
\hat{\Pcal} = (1/i\hbar) [H_{0},H^{\prime}_{\text{int}}] . 
\end{align}

As for the correlation $\Qcal(t)$, we cannot use the Eq.~\eqref{Entanglement} to this case because we have a tripartite system \cite{Meyer01,Brennen01,Filho01}. Therefore, one defines a quantity based on the average purity of each subsystem as $\Qcal_{\text{av}}\!=\!\sum_{n=1}^{N} \Qcal_{n}/N$ , where $\Qcal_{n}\!=\!\trs(\rho_{n}^2)$ with $\rho_{n}$ being the reduced matrix density of the $n$-th cell. We stress here that the above quantity cannot be taken as a measure of correlations for a general $\rho(t)$, but in the case where $\rho(t)$ is a pure state, it can be used as a measure of non-separability (correlations) of the system state. In fact, for a separable state of $N$ qubits $\trs(\rho_{n}^2)\!=\!1$, for all $N$, then we get $\Qcal_{\text{av}}\!=\!1$ for a fully uncorrelated state of $N$ qubits. Otherwise, in the case where the system is correlated (even for nearest-neighbor qubits) we shall find $\trs(\rho_{n}^2)\!\neq\!1$ for some $n$, then revealing a correlated system.

The Fig.~\ref{Fig4} shows the relevant quantities for the three-cell quantum battery. From Figs.~\eqref{Fig-Ergo2} and~\eqref{Fig-Ent2} it is possible to see that an entanglement charging process implies into a non-optimal charging process, since the case without correlation achieves maximum charge at $t\!=\!t_{\text{min}}$. Furthermore, we remark that this case corresponds to the situation where maximum coherence is created in the system during its evolution. Through a detailed analysis of the Fig.~\ref{Fig4}, it is not evident that we have a trivial trade-off between correlations and power, but if we take the coherence into account to a better understanding of the system.

\emph{Conclusions.} In this work, we studied the relation between entanglement and coherence with the performance of two- and three-cell quantum batteries. By using a system of coupled two-level systems we explore the role of an anisotropy parameter of the XXZ Heisenberg linear chain. Through a counterexample, we have shown that the generation of entanglement along the charging process of QBs can negatively contribute to the performance of QBs. Our results suggest that a non-trivial relation between the amount of entanglement and high-performance QBs is not universal and depends on the system we are dealing with. On the other hand, coherence develops a relevant role as the resource for efficiency of the system considered in our study. As a general conclusion, we highlight a correlation-coherence trade-off in the optimal performance of QBs, so that the high charging efficiency of the QBs adopted here cannot be explained by correlations only.

It is evident we recognize the validity of the large number of works in the literature showing the role of collective charging processes for scalable $N$-cell QBs. However, we highlight here the requirement of a detailed analysis of the real role of quantum correlations in the collective charging of such devices. By considering the results present in this paper and previous discussion on the work extraction from the coherence of quantum states~\cite{Kwon:18,Alexia:20,Baris:20}, we stress that a possible ``quantum supremacy" of QB needs to be investigated in more details. In addition, the definition of a class of different devices and charging processes would be a consequence of this study. The extension of this work to a scenario of $N$-cell QBs is content for future research, where a study of which physical quantity can be a good resource for optimal performance  of QBs can be appropriately provided.

\emph{Acknowledgments.} This work has been supported by the University of Kurdistan. F. T. Tabesh and S. Salimi thank Vice Chancellorship of Research and Technology,  University of Kurdistan. A. C. Santos acknowledges the financial support through the research grant from the São Paulo Research Foundation (FAPESP) (grant 2019/22685-1).

\bibliography{mybib-URL.bib}

\appendix

\section{Analytical solution for the system dynamics} \label{ApA}
The most general state of two qubits reads as
\begin{align}
\vert\Psi(0)\rangle=\mu\ket{\uparrow\uparrow} +\nu\ket{\uparrow\downarrow} +\eta\ket{\downarrow\uparrow} +\delta\ket{\downarrow\downarrow} .
\end{align}
with the help of Eqs. (\ref{6a}) and (\ref{7}), its time evolution will be
\begin{align}
\vert\Psi(t)\rangle=\mu(t)\ket{\uparrow\uparrow} +\nu(t)\ket{\uparrow\downarrow} +\eta(t)\ket{\downarrow\uparrow} +\delta(t)\ket{\downarrow\downarrow} ,
\end{align}
where
\begin{align}
\mu(t) &= -\frac{(\delta -\mu)}{2}e^{-iE_{1}t}+(\delta +\mu)(\gamma^{2}_{1}e^{-iE_{3}t}+\gamma^{2}_{2}e^{-iE_{4}t}) \nonumber\\
&+  \gamma_{1}\gamma_{2}(\nu +\eta)(e^{-iE_{4}t}-e^{-iE_{3}t}),\nonumber\\
\nu(t) &=-\frac{(\eta -\nu)}{2}e^{-iE_{2}t}+(\eta +\nu)(\gamma^{2}_{2}e^{-iE_{3}t}+\gamma^{2}_{1}e^{-iE_{4}t}) \nonumber\\
&+ \gamma_{1}\gamma_{2}(\delta +\mu)(e^{-iE_{4}t}-e^{-iE_{3}t}),\nonumber\\
\delta(t)&=(\delta -\mu)e^{-iE_{1}t}+\mu(t),\nonumber\\
\eta(t)&=(\eta -\nu)e^{-iE_{2}t}+\nu(t).
\end{align}
At this point, let's consider the most general state of two non-entangled qubits by
\begin{eqnarray}\label{11}
\mu &=& \sin[\theta_{1}]\sin[\theta_{2}]e^{i(\varphi_{1}+\varphi_{2})},\nonumber \\
\nu &=&\sin[\theta_{1}]\cos[\theta_{2}]e^{i\varphi_{1}},\nonumber\\
\eta &=&\cos[\theta_{1}]\sin[\theta_{2}]e^{i\varphi_{2}},\nonumber\\
\delta &=&\cos[\theta_{1}]\cos[\theta_{2}],
\end{eqnarray}
where we can consider $\theta_{1}, \theta_{2}\in [0,\pi]$ and $\varphi_{1}, \varphi_{2}\in [0,2\pi]$. Also, we find the energy $\tr(\rho(t)H_{0})$ 
\begin{eqnarray}
U(t) &=&-2\omega_{0}[\Gamma_{1}(\gamma^{2}_{1}\cos[(E_{3}-E_{1})t]+\gamma^{2}_{2}\cos[(E_{4}-E_{1})t])\nonumber\\
&+&\Gamma_{2}(\gamma^{2}_{1}\sin[(E_{3}-E_{1})t]+\gamma^{2}_{2}\sin[(E_{4}-E_{1})t])\nonumber\\
&+&\Gamma_{3}(\sin[(E_{4}-E_{1})t]-\sin[(E_{3}-E_{1})t])])\nonumber\\
&+&\Gamma_{4}(\cos[(E_{4}-E_{1})t]-\cos[(E_{3}-E_{1})t]),
\end{eqnarray}
with
\begin{align}\label{13}
\Gamma_{1}&=2(\cos^{2}[\theta_{1}]\cos^{2}[\theta_{2}]-\sin^{2}[\theta_{1}]\sin^{2}[\theta_{2}]),\nonumber\\
\Gamma_{2}&=(\sin[2\theta_{1}]\sin[2\theta_{2}]\sin[\varphi_{1}+\varphi_{2}]),\nonumber\\
\Gamma_{3}&=(\sin[2\theta_{1}]\sin[\varphi_{1}]+\sin[2\theta_{2}]\sin[\varphi_{2}]),\nonumber\\
\Gamma_{4}&=\gamma_{1}\gamma_{2}[(\sin[2\theta_{1}]\cos[2\theta_{2}]\cos[\varphi_{1}]\nonumber\\
&+\sin[2\theta_{2}]\cos[2\theta_{1}]\cos[\varphi_{2}])].
\end{align}
In addition, we define the instantaneous charge (ergotropy) as
\begin{align}\label{5}
\mathcal{E}(t)=U(t)-E_{\text{emp}},
\end{align}
and the instantaneous power
\begin{align}
\mathcal{P}(t)=\frac{d}{dt}\mathcal{E}(t).
\end{align}

At the beginning of the charging process, the battery is assumed to be empty, i.e., $\rho(0)=\vert emp\rangle\langle emp\vert$, this is achieved when we have $\theta_{1}=\theta_{2}=0$ in Eq. (\ref{11}), which leads to $\Gamma_{1}=2$ and $\Gamma_{2}=\Gamma_{3}=\Gamma_{4}=0$ in Eq. (\ref{13}).
Therefore we have
\begin{align}\label{16}
\mathcal{E}(t)&=-4\omega_{0}(\gamma^{2}_{1}\cos[(E_{3}-E_{1})t]+\gamma^{2}_{2}\cos[(E_{4}-E_{1})t]-\frac{1}{2}),
\end{align}
and 
\begin{align}\label{17}
\mathcal{P}(t)&=4\omega_{0}(\gamma^{2}_{1}(E_{3}-E_{1})\sin[(E_{3}-E_{1})t]\nonumber\\
&+\gamma^{2}_{2}(E_{4}-E_{1})\sin[(E_{4}-E_{1})t]).
\end{align}

\section{Dephasing effects on battery performance} \label{BpB}

In order to discuss the performance of the battery concerning decoherence and to describe how the coherence is a resource to the charging process, in this section we briefly present some results of the performance of a two-qubit QB driven by the Hamiltonian $H\!=\!H_{\text{ch}}+H_{\text{int}}$, where $H_{\text{ch}}\!=\!\hbar \Omega\sum_{n=1}^{2}\sigma^{x}_{n}$ and $H_{\text{int}}$ given by the Eq.~\eqref{Hint0}. Because we are interested in understanding the role of quantum coherence for the charging process, it is worthwhile to study battery performance in the presence of dephasing. For this purpose, we investigate our charging protocol driven by the Lindblad master equation for dephasing as
\begin{align}\label{LindEqDeph}
\frac{d\rho(t)}{dt}&=-\frac{i}{\hbar}[H,\rho(t)]+\gamma\sum_{i=1,2}\left(\sigma^{z}_{i}\rho(t)\sigma^{z}_{i}-\rho(t)\right),
\end{align}
with the local Lindblad operators $\sigma^{z}_{i}~(i=1,2)$ acting on each qubit with identical dephasing rate $\gamma$. 

\begin{figure}[t!]
	\subfloat[]{\includegraphics[scale=0.275]{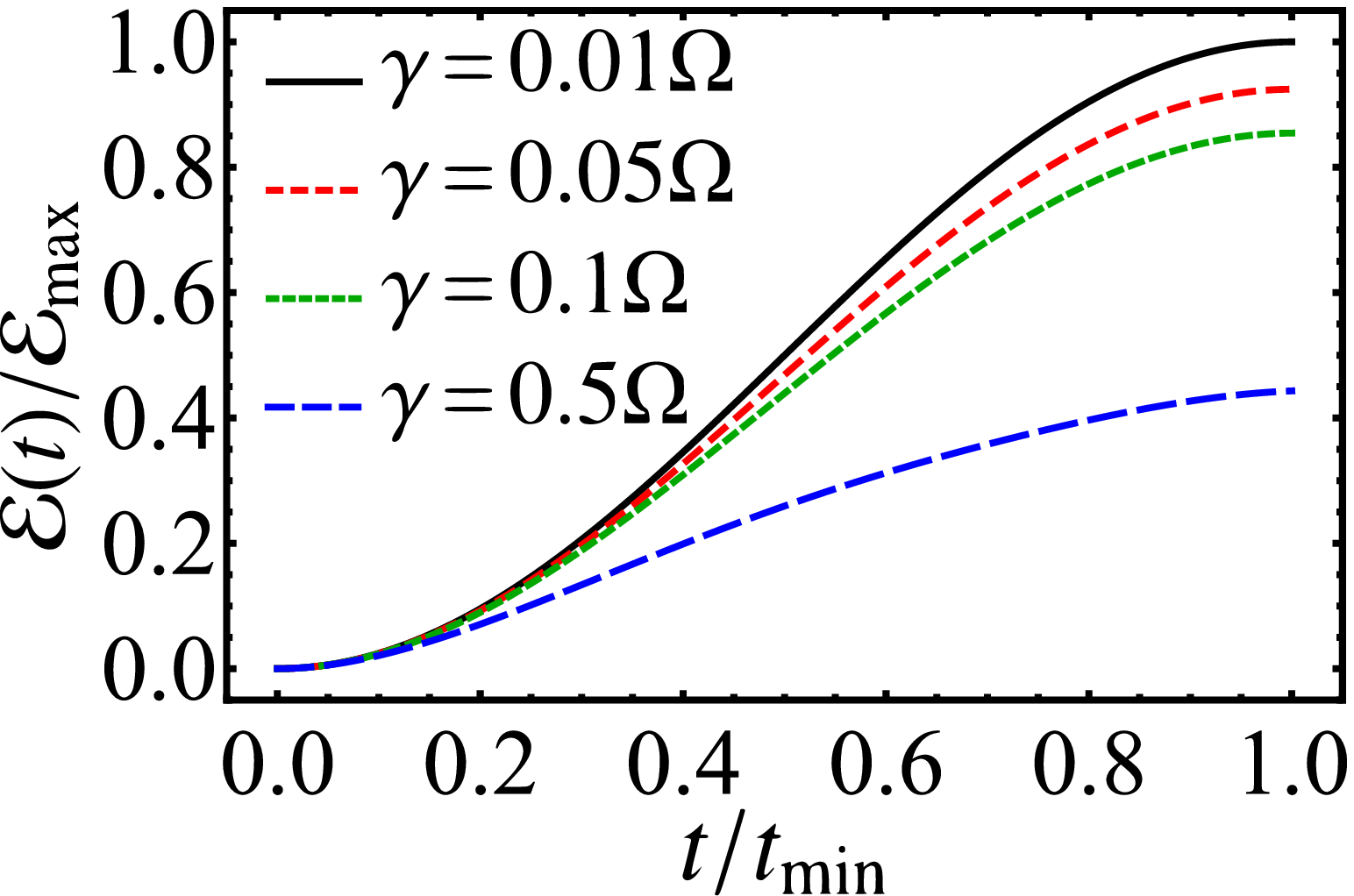}\label{Dec1}}~
	\subfloat[]{\includegraphics[scale=0.275]{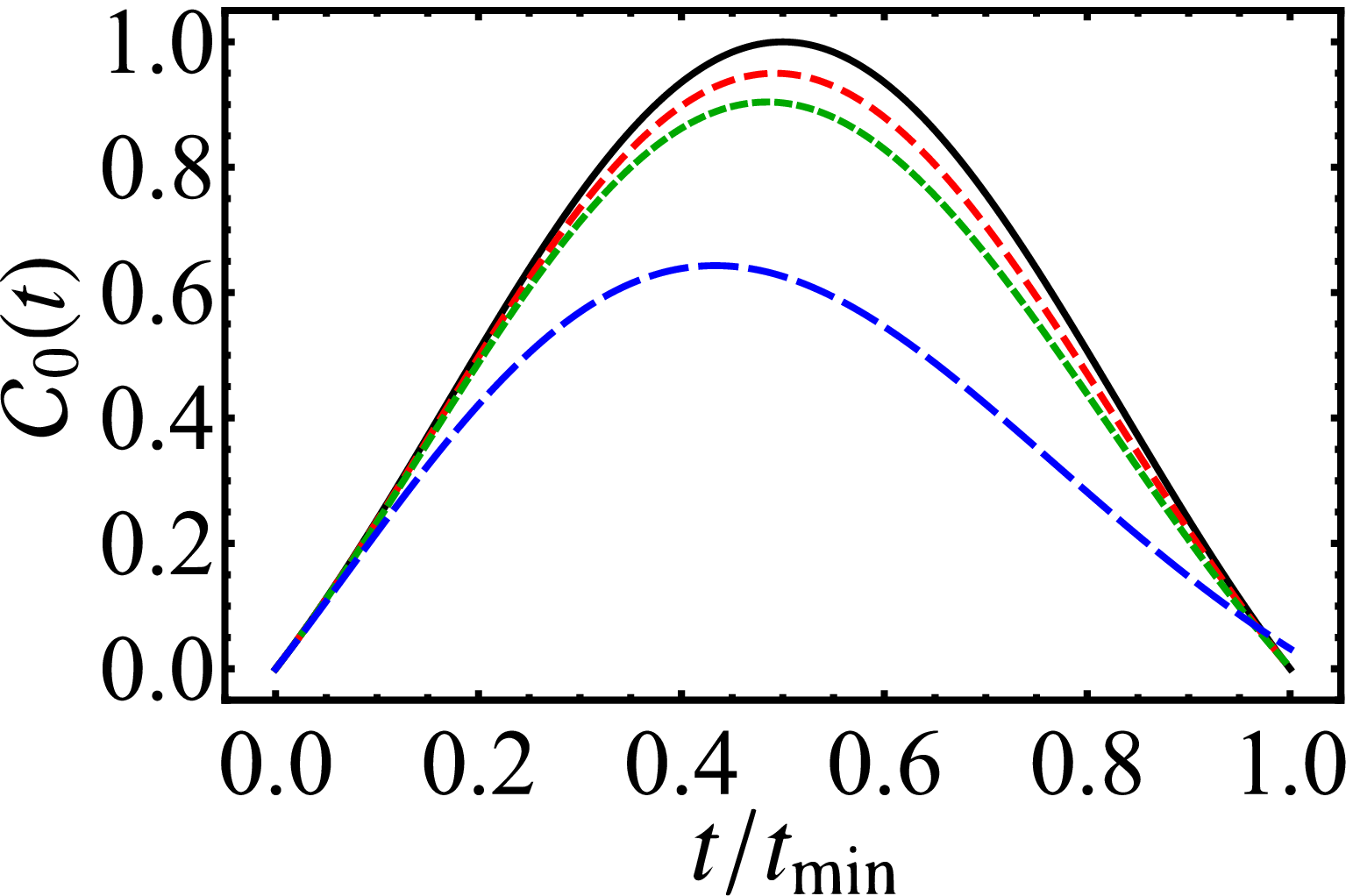}\label{Dec2}}
	\caption{\eqref{Dec1} Instantaneous ergotropy and \eqref{Dec2} coherence for the dissipative charging process of the two-cell QB driven by the Eq.~\eqref{LindEqDeph} for different values of the dissipative charging process. The Hamiltonian parameters are $J\!=\!\Omega$, $t_{\text{min}}\!=\!\pi/2\Omega$ and $\Delta\!=\!1$.} \label{DecoherenceGraphs}
\end{figure}

Now, because the dynamics is not unitary and leads the system to a non-pure density matrix, the ergotropy cannot be computed from internal energy, as we did in Eq.~\eqref{W}. In fact, for a general density matrix $\rho$ with dimension $N$ the ergotropy given in Eq.~\eqref{ErgGene} reads as~\cite{Allahverdyan:04,Baris:20}
\begin{align}
\Ecal = \sum\nolimits_{i,n}^{N,N} x_{n}\epsilon_{i} \left( |\interpro{x_{n}}{\epsilon_{i}}|^2 - \delta_{ni} \right) , \label{ErgoSimpl}
\end{align}
where $\ket{x_{n}}$ and $x_{n}$ are the eigenvectors and eigenvalues of $\rho$, so that $x_{1}\!\geq\!x_{2}\!\geq\!\cdots\!\geq\!x_{N}$, which obtained from the spectral decomposition of $\rho$, and $\epsilon_{i}$ are eigenvalues of the reference Hamiltonian $H_{0}$ with eigenstates $\ket{\epsilon_{i}}$, with $\epsilon_{1}\!\leq\!\epsilon_{2}\!\leq\!\cdots\!\leq\!\epsilon_{N}$. Therefore, in Fig.~\ref{DecoherenceGraphs} we present the ergotropy and coherence for different dissipative rates $\gamma$, for the case there we have the anisotropy parameter $\Delta\!=\!1$ and $t\!\in\![0,t_{min}]$, which leads to the optimal parallel charging process of the battery. As one can be seen, concerning the absence of dephasing effects, i.e. $\gamma\!=\!0$, by increasing $\gamma$ we have destructive impact on battery performance such that the extractable work becomes smaller. This occurs due to the amplification of dephasing effects in the charging process, as we can see from the graph for ergotropy and coherence in Figs.~\eqref{Dec1} and~\eqref{Dec2}, respectively. Consequently, the coherence is a character that plays the role of a quantum advantage in the QBs.

\end{document}